\documentclass{article}
\usepackage[cp1251]{inputenc}
\usepackage[english]{babel}
\usepackage{amssymb}
\begin{document}

\title{Towards an Axiomatic Formulation of Noncommutative Quantum Field Theory. II}

\renewcommand{\thefootnote}{\fnsymbol{footnote}}
\author{ M. Chaichian$^{a}$, \ M.\,N. Mnatsakanova$^{b}$,  \ and
 Yu.\,S. Vernov$^{c,}$\footnotemark[2]
\vspace*{3mm} \\
\it \small  $^a$Department of Physics, University of Helsinki, P.O. Box
64, Helsinki, Finland\\
\it \small $^b$Skobeltsyn Institute of Nuclear Physics, Lomonosov
Moscow State University, Moscow, Russia \\
\it \small $^c$Institute for Nuclear Research of the Russian Academy of Sciences, Moscow, Russia}
\footnotetext[2]{Professor Yuri Vernov passed away in Moscow on 27 May 2015.}
\renewcommand{\thefootnote}{\arabic{footnote}}

\date{ \ }
\maketitle

\vspace{5mm}

\begin{abstract}
{Classical results of the axiomatic quantum field theory, namely the
irreducibility of the set of field operators, Reeh and
Schlieder's theorems and generalized Haag's theorem, are proven
in $SO(1,1)$ invariant quantum field theory, of which
an important example is noncommutative
quantum field theory. New consequences of generalized Haag's theorem
are obtained in $SO(1,3)$ invariant theories. It has
been proven that the equality of four-point Wightman functions in
two theories leads to the equality of elastic scattering
amplitudes and thus the total cross-sections in these theories.}
\end{abstract}

\section{Introduction}
Quantum field theory (QFT) as a mathematically rigorous and
consistent theory was formulated in the framework of the axiomatic
approach in the works of Wightman, Jost, Bogoliubov, Haag and
others (\cite{SW}--\cite{Haag}).

Within the framework of this theory on the  basis of  most general
principles such as Poincar\'{e} invariance, local commutativity
and spectrality, a number of fundamental physical results, for
example, the CPT-theorem and the spin-statistics theorem  were
proven \cite{SW}--\cite{BLOT}.

Noncommutative quantum field theory (NC QFT) being one of the
generalizations of standard QFT has been intensively developed
during the past years (for reviews, see \cite{DN, Sz}). The idea
of such a generalization of QFT ascends to Heisenberg, and it was
initially developed in Snyder's work \cite{Snyder}. The present
development in this direction is connected with the construction
of noncommutative geometry \cite{Connes} and new physical
arguments in favour of such a generalization of QFT \cite{DFR}.
Essential interest in NC QFT is also due the fact that in some
cases it is a low-energy limit of string theory \cite{SeWi}. The
simplest and at the same time most studied version of
noncommutative field theory is based on the following
Heisenberg-like commutation relations between coordinates:
\begin{equation} \label{cr}
[ \hat{x}^{\mu}, \hat{x}^{\nu}] =i \,\theta^{\mu \nu},
\end{equation}
where $\theta^{\mu\nu}$ is a constant antisymmetric matrix.

It is known that the construction of NC QFT in a general case
($\theta^{0i} \neq 0$) meets serious difficulties with unitarity
and causality \cite{GM}--\cite{CNT}. For this reason the version
with $\theta^{0i} = 0$ (space-space noncommutativity), in which
there do not appear such difficulties, and which is a low-energy
limit of the string theory, draws special attention. Then there is
a system of coordinates, where the only nonvanishing components
of $\theta^{\mu\nu}$ are $\theta^{12} = -\theta^{21} \neq 0$ \cite{ABZ}.
Thus, when $\theta^{0i} = 0$, without loss of generality it is possible
to choose coordinates $x^0$ and $x^3$ as commutative and
coordinates $x^1$ and $x^2$ as noncommutative.

The relation (\ref{cr}) breaks the Lorentz invariance of the
theory, while the symmetry under the $SO(1,1) \otimes SO(2)$ subgroup of the Lorentz group survives \cite{AB}.
Translational invariance is still valid.  Below we shall consider
the theory to be $SO(1,1)$ invariant with respect to
coordinates $x^0$ and $x^3$. Besides these classical groups of
symmetry, in the paper \cite{CKNT} it was shown, that the
noncommutative field theory with the commutation relation
(\ref{cr}) of the coordinates, and built according to the
Weyl-Moyal correspondence, has also a quantum symmetry, i.e.
the twisted Poincar\'{e} invariance.

In the works \cite{AGM,CMNTV} the Wightman approach was
formulated for NC QFT. The CPT theorem and the spin-statistics
theorem were proven for scalar fields in the case $\theta^{0i} = 0$.

In \cite{AGM} it was proposed that Wightman functions in the
noncommutative case can be written down in the standard form
\begin{equation} \label{wf}
W\,(x_1,  \ldots, x_n) = \langle \, \Psi_0, \varphi (x_1)
 \ldots \varphi (x_n) \, \Psi_0 \, \rangle,
\end{equation}
where $\Psi_0$ is the vacuum state. However, unlike the
commutative case, these Wightman functions are only $SO(1,1)
\otimes SO(2)$ invariant. Actually, in \cite{AGM} the CPT
theorem has been proven in the commutative theory, where Lorentz
invariance is broken down to $SO(1,1) \otimes SO(2)$
symmetry, and as in the noncommutative theory it is necessary to use
the $\star$-product at least in coinciding points.

In \cite{CMNTV} (see also ref. \cite{Sz}) it was proposed that in the noncommutative case
the usual product of operators in the Wightman functions has to be
replaced by the Moyal-type product both in coinciding and
different points:
$$
\varphi (x_1) \star \cdots \star \varphi (x_n) = \prod_{a<b\leq n} \exp{\left
({\frac{i}{2} \, \theta^{\mu\nu} \, \frac{ {\partial}}{\partial x^{\mu}_a} \,
\frac{ {\partial}}{\partial x^{\nu}_b}} \right)} \,\varphi (x_1)  \ldots
\varphi (x_n),
$$
\begin{equation} \label{mprod}
a = 1,2, \ldots n - 1.
\end{equation}
Such a product of operators is compatible with the twisted
Poincar\'{e} invariance of the theory \cite{CPT}, and it also reflects
the natural physical assumption that noncommutativity should
change the product of operators not only in coinciding points but
also in different ones. This follows also from another
interpretation of NC QFT in terms of a quantum shift operator
\cite{CNT05}.

In \cite{VM05} it was shown that in the derivation of axiomatic
results, the concrete type of product of operators in various
points is insignificant. It is essential only that from the
appropriate spectral condition (see formula (\ref{imp})), the
analyticity of Wightman functions with respect to the commutative
variables $x^0$ and $x^3$ follows, while $x^1$ and $x^2$ remain
real. In accordance with eq. (\ref{mprod}) the Wightman functions
can be written down as follows:
\begin{equation} \label{wfg}
W_{\star}\,(x_1,  \ldots, x_n) = \langle \Psi_0, \varphi
\,(x_1)\,\star\, \cdots \,\star\,  \varphi\,(x_n)  \Psi_0 \rangle.
\end{equation}

Note that actually there is no field operator defined in a point
\cite{Wight}, (see also \cite{BLT}). Only the smoothed operators
written symbolically as
\begin{equation} \label{vff}
\varphi_f \equiv\int \,\varphi\,(x) \,f\,(x) \, d \, x,
\end{equation}
where $f\,(x)$ are test functions, can be rigorously defined.

In QFT the standard assumption is that all $f\,(x)$ are test
functions of tempered distributions. On the contrary, in the NC
QFT the corresponding generalized functions can not be tempered
distributions, since the $\star$-product contains an infinite number of
derivatives. It is well-known (see, for example, \cite{SW}) that
there could be only a finite number of derivatives in any tempered
distribution.

The formal expression (\ref{wfg}) actually means that the scalar
product of the vectors $\Phi_{k} = \varphi_{f_k}\,  \cdots
\,\varphi_{f_1} \, \Psi_0$ and $\Psi_n = \varphi_{f_{k + 1}}\,
\cdots  \,\varphi_{f_n} \, \Psi_0$ is the following:
\begin{eqnarray} \label{scprod}
{\langle \, \Phi_{k}, \Psi_{n-k} \, \rangle }
 \nonumber  \\ =
\int\, W\,(x_1,  \ldots, x_n) \, \overline{f_1\,(x_1)}\,\star
\cdots \star \overline{f_k\,(x_k)} \star f_{k + 1}\,(x_{k + 1})\,\star\,  \cdots \star \,{f_n\,(x_n)}  d\,x_1 \ldots d\,x_n;
\nonumber  \\  W\,(x_1,  \ldots, x_n)
=\langle\Psi_0,\varphi(x_1)\cdots\varphi(x_n)\Psi_0\rangle.
\end{eqnarray}
It was shown in \cite{CMTV7} that the series
\begin{equation} \label{fsprodxy}
f\,(x) \star f\,(y) = \exp{\left ({\frac{i}{2} \, \theta^{\mu\nu}
\, \frac{ {\partial}}{\partial x^{\mu}} \, \frac{
{\partial}}{\partial y^{\nu}}} \right)} \,f\,(x)  f\,(y)
\end{equation}
converges if $f\,(x) \in S^\beta, \; \beta <  1 / 2$, where $S^\beta$ is
a Gel'fand-Shilov space \cite{GSh}. The same result was also
obtained in \cite{Sol07}.

The difference of the noncommutative case from the commutative one is that action of the operator $\varphi_f$ is defined by the $\star$-product.

In \cite{VM05} it was shown that, besides the above-mentioned
theorems, in NC QFT (with $\theta^{0i} = 0$) a number of other
classical results of the axiomatic theory remain valid. In \cite{CPT}
the Haag's theorem \cite{HaagTh, FD} (see
also \cite{SW} and references therein) was obtained on the basis
of the twisted Poincar\'{e} invariance of the theory.

The present work is a continuation and the completion of our previous work published in \cite{CMNTV}, and deals with further
development of the axiomatic approach in NC QFT.
In fact, our results are valid for a wide class of  $SO(1,1)$ invariant
four-dimensional field theories.

At first we formulate the basic properties of Wightman functions
in space-space NC QFT.

In the present work, analogues of some known results of the
axiomatic approach in quantum field theory are obtained for the $SO(1,1)$ invariant field theory, of which an important
example is NC QFT. We  prove that classical results, such as the
irreducibility of the set of field operators, the theorems of Reeh
and Schlieder \cite{SW}--\cite{BLOT} remain valid in the
noncommutative case. It should be emphasized that the results
obtained in this paper do not depend on the $SO(2)$
invariance of the theory in the variables $x^1$ and $x^2$ and
therefore can be extended to more general cases. The
irreducibility of the set of field operators remain valid in any
theory, which is translation invariant in commutative variables,
if only eq. (\ref{26}) is fulfilled. The first theorem of Reeh and
Schlieder is valid, if the Wightman functions are analytical in
the variables $x^0$ and $x^3$ in the primitive domains of
analyticity ("tubes").

In the $SO(1,3)$  invariant theory new consequences of the
generalized Haag's theorem are found, without analogues in NC QFT.
At the same time it is proven that the basic physical conclusion
of Haag's theorem is valid also in the $SO(1,1)$ invariant
theory, and it is sufficient that spectrality, local commutativity
condition and translational invariance be fulfilled only for the
transformations concerning the commutating coordinates. The
analysis of Haag's theorem reveals essential distinctions between
commutative and noncommutative cases, more precisely between the
$SO(1,3)$ and $SO(1,1)$ invariant theories. In the
commutative case, the conditions (\ref{con1}) and (\ref{con2}),
whose consequence is generalized Haag's theorem, lead to the
equality  of Wightman functions in two theories up to four-point
ones. In the present paper it is shown that in the $SO(1,
1)$ invariant theory, unlike the commutative case, only two-point
Wightman functions are equal, and it is shown that from the
equality of two-point Wightman functions in two theories, it
follows that if in one of them the current is equal to zero, it is
equal to zero in the other as well and under weaker conditions
than the standard ones. It is also shown that for the derivation
of eq. (\ref{con2}) it is sufficient to assume that the vacuum
vector is a unique normalized vector, invariant under translations
along the axis $x^3$.  It is proven that from the equality of
four-point Wightman functions in two theories, the equality of
their elastic scattering amplitudes follows and, owing to the
optical theorem, the equality of total cross sections as well. In the
derivation of this result, the condition of local commutativity (LCC)
is not used.

The study of Wightman functions leads still to new nontrivial
consequences also in the commutative case\footnote{A partial
result on the subject had been previously communicated in
\cite{CMTV}.}.

The paper is arranged as follows. In section 2 the basic
properties of Wightman functions in space-space NC QFT are
formulated; in section 3 the irreducibility of the set of field
operators is proven; in section 4 generalizations of the theorems
of Reeh and Schlieder to NC QFT are obtained; section 5 is devoted
to generalized Haag's theorem; in section 6 it is shown that in
the commutative case, the conditions of weak local commutativity
(WLCC) and of local commutativity (LCC), which are valid in the
noncommutative case ((\ref{wftx}) and (\ref{wftxi})), appear to be
equivalent to the usual WLCC and LCC, respectively.

\section{Basic Properties of Wightman Functions in Space-space NC QFT }

As in the commutative case, we assume that every vector from
the space of the complete set of all physical states,  $J$,
can be approximated with arbitrary accuracy by the vectors of the type
type
\begin{equation}\label{vc}
\varphi_{f_1}\, \cdots \,\varphi_{f_n}\,\Psi_{0}.
\end{equation}
In other words, the vacuum vector $\Psi_{0}$ is cyclic, i.e. the
axiom of cyclicity of vacuum is fulfilled.

Let us note that the vectors of the type (\ref{vc}) can be written
formally as follows:
\begin{equation}\label{vcfor}
\varphi_{f_1}\, \cdots \,\varphi_{f_n}\,\Psi_{0} =
\int\,\varphi(x_1)\cdots\varphi(x_n)\Psi_0 \, f_1\,(x_1)\,\star
\cdots \star \,{f_n\,(x_n)}\, d\,x_1 \ldots d\,x_n.
\end{equation}
It is natural to assume that Wightman functions are tempered
distributions with respect to commutative coordinates as the
$\star$-product contains derivatives with respect to
noncommutative coordinates only. In accordance with this
assumption we can use the standard arguments to prove Wightman
functions analyticity in "tubes" and extended "tubes".

It is well known that in commutative case the analyticity of Wightman functions
in tubes is a consequence of the spectral condition,
which implies that the complete system of physical states (in gauge
theories also nonphysical ones) does not contain tachyon states in
momentum space. It means that momentum $P_m$ for every state
satisfies the condition: $P_n^0 \geq | \vec{P_n} |$. This
condition is usually written as $P_n \in \bar{V^+}$. Since Wightman
functions in the noncommutative case are analytical function only
in commutative variables, it is sufficient to assume the weaker
condition of spectrality. Precisely, we assume that any vector in
$p$ space, belonging to the complete system of these vectors, is
time-like with respect to momentum components  $P_n^0$ and
$P_n^3$, i.e. that
\begin{equation} \label{imp}
P_n^0 \geq | P_n^3 |.
\end{equation}
The condition (\ref{imp}) is conveniently written as $P_n \in
\bar{V_2^+}$, where $\bar{V_2^+}$ is the set of the
four-dimensional vectors satisfying the condition $P^0 \geq |P^3
|$.

For the results obtained below, translational invariance only in
commuting coordinates is essential, therefore we write down the
Wightman functions as:
\begin{equation} \label{ewf}
W\,(x_1,  \ldots, x_n) = W\,(\xi_1,   \ldots, \xi_{n - 1}, X),
\end{equation}
where $X$ designates the set of  noncommutative  variables $x_i^1,
x_i^2, \;  i = 1, \ldots n$, and commutative  variables   $\xi_j  = \{\xi_j^0, \xi_j^3 \}$,
where $\xi_j^0 = x_j^0 - x_{j +1}^0, \xi_j^3 = x_j^3 - x_{j
+1}^3$, $j=1,...,n-1$.

Thus at arbitrary $X$ we can express the scalar product (\ref{scprod}) as follows:
\begin{equation}\label{scprod_2}
\langle \, \Phi_{k}, \Psi_n \, \rangle = \int W\,(\xi_1,   \ldots,
\xi_{n - 1}, X) \, f\,(\xi_1,   \ldots, \xi_{n - 1}, X) \,
d\,\xi_1 \ldots d\,\xi_{n - 1},
\end{equation}
where $ f\,(\xi_1,   \ldots, \xi_{n - 1}, X)$ given in  (\ref{scprod}),
and use the completeness of the system of vectors $\Psi_{P_m}$, where $P_m =
\{ P_n^0, P_n^t  \}$ is the two-dimensional momentum corresponding to the
commutative coordinates, multiindex $n$ denotes all other characteristics of
the state. So
\begin{equation} \label{last}
\langle\Phi, \Psi\rangle = \sum_n\int d \,P_m \langle\Phi, \Psi _{P_m} \rangle
\langle\Psi_{P_m}, \Psi\rangle.
\end{equation}
From the condition (\ref{imp}) and eq. (\ref{last}) it follows
that
\begin{equation} \label{10}
\int \, d \, a \, e^{- i \, p \, a} \, \langle\Phi, U \, (a) \,
\Psi\rangle = 0, \quad \mbox{if} \quad p \not \in \bar V_2^+,
\end{equation}
where $a = \{a^0, a^3 \}$ is a two-dimensional vector, $U \, (a)$ is a
translation in the plane $x^0, x^3$, and $\Phi$ and $\Psi$ are arbitrary
vectors. The equality (\ref{10}) is similar to the corresponding equality in
the standard case (\cite{SW}, Chap.~2.6).

A direct consequence of the equality (\ref{10}) is the spectral
property of Wightman functions:
\begin{equation} \label{wfsp}
W\, (P_1..., P_{n - 1}, X) = \frac{1}{(2\pi)^{(n-1)/2}} \, \int \,
e^{i \, P_j \,\xi_j} \, W \, (\xi_1..., \xi_{n - 1}, X) \, d
\,\xi_1 ... d \,\xi_{n - 1} = 0,
\end{equation}
if $P_j \not \in \bar V_2^+$. The proof of the equality (\ref{wfsp}) is similar to the proof of the spectral condition in
the commutative case \cite{SW}, \cite{BLOT}. Recall that in the
latter case the equality (\ref{wfsp}) is valid, if $P_j \not \in
\bar V^+$. Having written down $W\,(\xi_1,   \ldots, \xi_{n - 1},
X)$ as
\begin{equation} \label{wfspo}
W\, (\xi_1..., \xi_{n - 1}, X) = \frac{1}{(2\pi)^{(n-1)/2}}  \,
\int \, e^{- i \, P_j \,\xi_j} \, W \, (P_1..., P_{n - 1}, X) \, d
\, P_1 ... d \, P_{n - 1},
\end{equation}
and taking into  account that Wightman functions are tempered
distributions with respect to the commutative variables, we obtain
that, due to the condition (\ref{wfsp}), $W\,(\nu_1,   \ldots,
\nu_{n - 1}, X)$ is analytical in the "tube" $\; T_n^-$:
\begin{equation} \label{tube}
{\nu}_i \in T_n^-, \quad \mbox{if} \quad {\nu}_i = {\xi}_i - i \,{\eta}_i, \;
{\eta}_i \in V_2^+, \; {\eta}_i = \{{\eta}_i^0, {\eta}_i^3 \}.
\end{equation}
It should be stressed that the noncommutative coordinates $x_i^1,
\; x_i^2$ remain always real.

Owing to  $SO(1,1)$ invariance and according to the
Bargmann-Hall-Wightman theorem \cite{SW}--\cite{BLOT},
$W\,(\nu_1,   \ldots, \nu_{n - 1}, X)$ is analytical in the domain
$T_n$,
\begin{equation} \label{ctn}
T_n = \cup\, {\Lambda}_c \,T_n^-,
\end{equation}
where ${\Lambda}_c \in SO_c \, (1, 1)$ is the two-dimensional
analogue of the complex Lorentz group. This expansion is similar
to the transition from tubes to expanded tubes in the commutative
case.  Just as in the commutative case, the expanded domain of
analyticity contains real points $x_{i}$, which are the
noncommutative Jost points, satisfying the condition $x_{i} \sim
x_{j}, \; \forall \; i, j$, which means that
\begin{equation} \label{app}
{ \left (x_i^0 - x_j^0\right)}^2 - {\left (x_i^3 - x_j^3\right)}^2
< 0.
\end{equation}
It should be emphasized that the noncommutative Jost points are a
subset of the set of Jost points of the commutative case, when
\begin{equation} \label{pp}
{ \left (x_i - x_j\right)}^2 < 0 \qquad \forall \; i, j.
\end{equation}
Let us proceed to the LCC in space-space NC QFT.

First let us recall this condition in commutative case. In the
operator form this condition is
\begin{equation} \label{vffod}
[\varphi_{f_1}, \varphi_{f_2}] = 0, \quad \mbox{if} \quad O_{1}\sim O_{2},
\end{equation}
where $O_{1} = \mbox{supp} \, f_1$, $O_{2} = \mbox{supp} \, f_2$.
The condition $O_{1}\sim O_{2}$ means that ${(x - y)}^2 < 0 \:
\forall \, x \,\in \, O_{1}$ and $ y \,\in \, O_{2}$. The
condition (\ref{vffod}) is equivalent to the following property of
Wightman functions:
\begin{equation} \label{wftxi}
W\,(x_1,   \ldots, x_i,  x_{i +1}, \ldots,  x_n) = W\,(x_1,
\ldots, x_{i + 1}, x_i, \ldots,  x_n),
\end{equation}
if supp $f_i \in O_i $, supp $f_{i+1} \in  O_{i+ 1},  \; O_i \sim
O_{i+ 1}$.

In the noncommutative case we have the similar condition, but now
$O_{1}\sim O_{2}$ means that
$$
{(x^{0} - y^{0})}^{2} - {(x^{3} - y^{3})}^{2} < 0, \quad \forall
\, x \in O_{1}, \; y \in O_{2}.
$$
In terms of Wightman functions this condition means that
$$
\int W\,(x_1,   \ldots, x_i,  x_{i +1}, \ldots,  x_n) \,f\,(x_{1})
\star \cdots \star\, f\,(x_{i})\,  \star \,f\,(x_{i + 1}) \, \star
\cdots \star\, f\,(x_{n}) \, d\,x_1 \,\ldots \,d\,x_n =
$$\begin{equation}
 \int W\,(x_1,   \ldots, x_{i + 1}, x_i, \ldots,  x_n)\, f\,(x_{1})  \,\star \cdots \,
\star \,f\,(x_{i + 1}) \,\star \,f\,(x_{i}) \,\star \cdots \,\star
\,f\,(x_{n})\, d\,x_1 \ldots d\,x_n,
\end{equation}
where $W\,(x_1,  \ldots, x_n) \equiv \langle \Psi_0,
\varphi\,(x_1) \, \ldots \, \varphi\,(x_n) \, \Psi_0 \rangle$.

Let us point out that in the noncommutative case WLCC
\begin{equation} \label{wftx}
W\,(x_1,  \ldots, x_n) = W\,(x_n,  \ldots,  x_1), \quad \mbox{if}
\quad x_i \sim x_j \quad \forall \; i, j.
\end{equation}
has the same form as in the local theory with the same difference
as for LCC.

\section{Irreducibility of the set of field operators $\varphi_f$ in NC QFT}

The irreducibility of a set of field operators $\varphi_f$ implies that, from
the condition
\begin{equation} \label{24}
A \, \varphi_{f_1}\, \cdots  \,\varphi_{f_n}  \, \Psi_0 =
\varphi_{f_1}\, \cdots  \,\varphi_{f_n} \, A \,  \Psi_0
\end{equation}
where $f_i = f_i\,(x_{i})$ are arbitrary test functions and $A$ is
a bounded operator, follows that
\begin{equation} \label{25}
A = C \,\mathbb I , \qquad C \in \mathbb C ,
\end{equation}
where $\mathbb I$ is the identity operator.

In the noncommutative case the condition of irreducibility of the
set of operators $\varphi_f$ is  valid as well as in the commutative
case. The point is that for this it is sufficient to have the
translational invariance in the variable $x^0$ and the spectral
condition, which can be weakened up to the condition
\begin{equation} \label{26}
P_n^0 \geq 0.
\end{equation}
Using condition (\ref{24}) and the invariance of the vacuum vector
with respect to the translations $U \, (a)$ on the axis $x^0$, we
obtain the following chain of equalities
$$
\langle \, A^* \,\Psi_0, U \,(a) \, \varphi_{f_1}\, \cdots
\,\varphi_{f_n} \,\Psi_0 \,\rangle =
$$
$$
\langle \, \Psi_0, A \,U \, (a) \, \varphi_{f_1}\, \cdots
\,\varphi_{f_n} \,\Psi_0 \, \,\rangle = \langle \, \Psi_0,
\varphi_{f_{1}\,(x_{1} + a)}\, \ldots \,\varphi_{f_{n}\,(x_{n} +
a)}\,A\,\Psi_0 \,\rangle =
$$
$$
\langle \,\varphi_{\bar{f}_{n}\,(x_{n} + a)}\, \ldots
\,\varphi_{\bar{f}_{1}\,(x_{1} + a)}\,\Psi_0, A\,\Psi_0 \,\rangle =
$$
$$
\langle \,U\,(- a)\,\varphi_{\bar{f}_{n}\,(x_{n} + a)}\, \ldots
\,\varphi_{\bar{f}_{1}\,(x_{1} + a)}\,\Psi_0, U\,(- a)\,A\,\Psi_0 \,\rangle =
$$
\begin{equation} \label{27}
\langle \, \varphi_{\bar{f_n}}\, \cdots  \,\varphi_{\bar{f_1}}
\,\Psi_0, U \, (-a) \, A \,\Psi_0 \,\rangle.
\end{equation}
So
\begin{equation} \label{27a}
\langle \, A^* \,\Psi_0, U \, (a) \, \varphi_{f_1}\, \cdots
\,\varphi_{f_n} \,\Psi_0 \,\rangle = \langle \,
\varphi_{\bar{f_n}}\, \cdots  \,\varphi_{\bar{f_1}} \,\Psi_0, U \,
(-a) \, A \,\Psi_0 \,\rangle.
\end{equation}
In accordance with eq. (\ref{10})
$$
\int \, d \, a \, e ^{- i \, p^0 \, a} \, \langle \, A ^*
\,\Psi_0, U\, (a) \, \varphi_{f_1}\, \cdots  \,\varphi_{f_n}
\,\Psi_0 \,\rangle \neq 0,
$$
only if $p^0 \geq 0$. However,
$$
\int \, d \, a \, e ^{- i \, p^0 \, a} \, \langle
\,\varphi_{\bar{f_n}}\, \cdots  \,\varphi_{\bar{f_1}} \,\Psi_0, U
\, (- a) \, A \,\Psi_0 \,\rangle \neq 0,
$$
only if $p^0 \leq 0$. Hence, the equality (\ref{27}) can be
fulfilled only when $ p^0 = 0$. As we assume the absence of
vectors noncollinear to the vacuum one and satisfying the
condition $P^{0} = 0$,  there is no vector distinct from the
vacuum one, which contributes to both left and right parts of eq.
(\ref{27}) simultaneously. Taking into account the completeness of
the system of vectors $\Psi_{P_{n}}$ we come to conclusion that
\begin{equation} \label{28}
A\,\Psi_0 = C \,\Psi_0,
\end{equation}
as $\varphi_{f_1}\, \cdots  \,\varphi_{f_n}  \,\Psi_0$ is an
arbitrary vector. Thus owing to (\ref{24}) and (\ref{28})
\begin{equation} \label{29}
A\,\varphi_{f_1}\, \cdots  \,\varphi_{f_n}  \,\Psi_0 = C
\,\varphi_{f_1}\, \cdots  \,\varphi_{f_n} \,\Psi_0.
\end{equation}
The required equality (\ref{25}) follows from eq. (\ref{29}) in
accordance with the boundedness  of the operator $A$ and cyclicity
of the vacuum vector.

\section{Cluster properties and their consequences}

 It is known \cite{SW, BLT} that in commutative theory Wightman
functions satisfy the following cluster properties:
\begin{eqnarray}\label{clusprop}
W\,(x_{1}, \ldots x_{k}, x_{k + 1} + \lambda a, \ldots x_{n} +
\lambda a) \to W\,(x_{1}, \ldots x_{k})\, W\,(x_{k + 1}, \ldots
x_{n}),
\end{eqnarray}
if $\lambda \to \infty $  and  $a^{2} = - 1$.
Let us show how the classical proof (see \cite{SW}) can be extended to space-space NC QFT.

First let us point out that in the commutative case the translation vector can be arbitrary, but in noncommutative case this vector has to belong to the commutative plane.
Surely in the commutative case we can also choose the translation vector to be in this plane. If we do this the proof in NC QFT is similar to the corresponding proof in usual QFT.
As in  \cite{SW}  we give the proof for theories with a mass gap.
In commutative case we use the following properties of Wightman functions:
\begin{enumerate}
    \item [i]  corresponding Wightman functions are tempered distributions;

   \item [ii]  LCC is valid.
\end{enumerate}
But if in the commutative case we make a shift  in the plane, which in the noncommutative case is a commutative plane, then LCC coincide in the commutative and noncommutative cases. Let us stress that it is sufficient to do translation in only one direction
as the translation vector is not in the final result. Taking into account that corresponding test functions in noncommutative case are tempered distributions in respect with commutative variables, we see that two crucial points in the derivation of cluster properties coincide in the commutative  and noncommutative cases in above-mentioned case of choosing translation vector.

Eq. (\ref{clusprop}) can be refined (see \cite{SW}). Namely,  if we consider the theory, where only massive particles exist,  in addition to the eq. (\ref{clusprop}) we have:
\begin{eqnarray}\label{cluspropref}
\left|W\,(x_{1}, \ldots x_{k}, x_{k + 1} + \lambda a, \ldots x_{n} +
\lambda a)  -  W\,(x_{1}, \ldots x_{k})\, W\,(x_{k + 1}, \ldots
x_{n})\right| < \frac{C}{\lambda^{n}},
\end{eqnarray}
where $n$ is arbitrary.
If the theory contains a massless particle, then in inequality (\ref{cluspropref}) $n \leq 2$.
The first case corresponds to the theories with short-range interaction, the second vase to long-range ones.
For Coulomb law $n = 2$ in  inequality (\ref{cluspropref}) \cite{St}.

Let us pass to the proof. \\
We consider two functions:
\begin{eqnarray}\label{efone}
F_{1} = W\,(x_{1}, \ldots, x_{k}, x_{k + 1} + \lambda a, \ldots, x_{n} +
\lambda a) - W\,(x_{1}, \ldots, x_{k})\, W\,(x_{k + 1}, \ldots,
x_{n})
\end{eqnarray}
and
\begin{eqnarray}\label{eftwo}
F_{2} = W\,( x_{k + 1} + \lambda a, \ldots, x_{n} +
\lambda a, x_{1}, \ldots, x_{k}) - W\,(x_{1}, \ldots, x_{k})\, W\,(x_{k + 1}, \ldots,
x_{n}).
\end{eqnarray}
If $\lambda \to \infty $  and  $a^{2} = - 1$ and $a \in \{x^{0}, x^{3}\}$, then owing  LCC,  in space-space NC QFT
\begin{equation}\label{equal}
F_{1} = F_{2}.
\end{equation}
 The simplest choice is:  $a = \{0, 1\}$   and  it would be our choice.
 It is easy to see that $ F_{1} = F_{2} = 0$   at any $\lambda$  if $P^{2} < M^{2}$   as we consider theories with a mass gap.
 Indeed let us put the complete system of vectors  $ \Psi_{P, n}$   between  points $x_{k}$  and   $x_{k + 1}$. Then we have: 
\begin{equation} \label{wfvec}
 \sum\limits_n\,\int\,d\,P\, \langle \Psi_0, \varphi \,(x_1)\, \cdots \,  \varphi\,(x_k)  \Psi_{P, n} \rangle \, \langle \Psi_{P, n},  \varphi\,(x_{k + 1} + \lambda\,a) \,  \cdots \,  \varphi\,(x_{n} + \lambda\,a)   \Psi_0 \rangle,
 \end{equation}
where  $ n $ denotes all other quantum numbers. Then we have
\begin{equation} \label{wfu}
 \sum\limits_n\,\int d\,P\, \langle \Psi_0, \varphi \,(x_1)\, \cdots \,  \varphi\,(x_k)  \Psi_{P, n} \rangle \, \langle \Psi_{P, n}, \,U\,(\lambda\,a)\, \varphi\,(x_{k + 1}) \,  \cdots \,  \varphi\,(x_{n})   \Psi_0 \rangle,
 \end{equation}
 where $U\,(a)$ is a translation operator.  Let us recall that  $U\,(a)\, \Psi_0 =  \Psi_0$. Then
$$
 \sum\limits_n\,\int\,d\,P\, \langle \Psi_0, \varphi \,(x_1)\, \cdots \,  \varphi\,(x_k)\, \Psi_{P, n} \rangle \,   \langle U\,(- \lambda\,a)  \Psi_{P, n}, \,\varphi\,(x_{k + 1}) \,  \cdots \,  \varphi\,(x_{n})   \Psi_0 \rangle =
 $$
 \begin{equation} \label{wfexp}
 \sum\limits_n\,\int\,d\,P\, \exp\,(-i\,\lambda\,a\, P) \, \langle \Psi_0, \varphi \,(x_1)\, \cdots \,  \varphi\,(x_k)\, \Psi_{P, n} \rangle \, \langle \Psi_{P, n}, \,\varphi\,(x_{k + 1}) \,  \cdots \,  \varphi\,(x_{n})   \Psi_0 \rangle .
 \end{equation}
 Thus using a translation along axis $x_{3}$ as before, we see that
 $$
 F_{1} =
 $$
 $$
 \sum\limits_n\int d\,P\, \exp\,(-i\,\lambda\, P_{3}) \, \langle \Psi_0, \varphi \,(x_1)\, \cdots \,  \varphi\,(x_k)\, \Psi_{P, n} \rangle \, \langle \Psi_{P, n}, \,\varphi\,(x_{k + 1}) \,  \cdots \,  \varphi\,(x_{n})   \Psi_0 \rangle   -
 $$
 \begin{equation} \label{efexp}
 W\,(x_{1}, \ldots, x_{k})\, W\,(x_{k + 1}, \ldots, x_{n}).
 \end{equation}
 As  $P_{3} = 0$ for $\Psi_{0}$, we see that  $F_{1} \neq 0$ only if   $P^{2} \geq M^{2}$.
 The same is true for function  $F_{2}$.

 Now let us take into account that Wightman functions in space-space NC QFT are tempered distributions with respect to the commutative coordinates.  It means that
\begin{eqnarray} \label{tempdistr}
\int F\,(x_{1}, \ldots, x_{n}) h\,(x_{1}, \ldots, x_{n})\,d\,x_{1} \ldots d\,x_{n}   \nonumber  \\   =
\int (D^{m}\,G)\,(\lambda, x_{1}, \ldots, x_{n})  h\,(x_{1}, \ldots, x_{n})\,d\,x_{1} \ldots d\,x_{n},
\end{eqnarray}
where  $ h\,(x_{1}, \ldots, x_{n})$ is a test function and  $F = F_{1} - F_{2}$.  As $ F_{1} - F_{2} = 0$ at $\lambda \to \infty$, then  $(D^{m}\,G) = 0$  if $\lambda \to \infty$.
Let us show that actually
\begin{equation} \label{de}
 (D^{m}\,G)\,(\lambda, x_{1}, \ldots, x_{n})  = 0,
\end{equation}
if $R^{2} < R_{0}^{2}$, where $R^{2} = \sum\limits_{j=1}^{n}[{(x_{j}^{0})}^{2}
+ {(x_{j}^{3})}^{2}]$ and $R_{0} = {1\over 4} \lambda$. \\
Indeed,
$$
{(x_{i} - x_{k} - \lambda\, a)}^{2}  = {(x_{i}^{0} - x_{k}^{0})}^{2} - {(x_{i}^{3} - x_{k}^{3})}^{2} - \lambda^{2} - 2 \lambda (x_{i}^{3} - x_{k}^{3}) \leq
$$
$$
2\,{\left(({x_{i}^{0})}^{2} + {(x_{k}^{0})}^{2}\right)} +   2 \lambda\,{\left(|x_{i}^{3}| + |x_{k}^{3}|\right)} - \lambda^{2}
\leq 2 R^{2}  + 2\lambda\,R - {\lambda}^{2} < 0
$$
\begin{equation} \label{uneq}
\mbox{if, for example,} \quad R_{0} = {\frac{1}{4}} \lambda \quad
\mbox{at} \quad \lambda \to \infty.
\end{equation}
So
\begin{equation} \label{ef}
F = \int\limits_{R_{0}} D^{m}\,G\,(\lambda, x_{1}, \ldots, x_{n})  h\,(x_{1}, \ldots, x_{n})\,d\,x_{1} \ldots d\,x_{n}.
\end{equation}
As $R_{0} \to \infty$  at  $ \lambda \to \infty$, and the integral in question converges, then $F \to 0$  at $ \lambda \to \infty$.
In order to see that also  $F_{1} \to 0$  at $\lambda \to \infty$, let us exchange  $ h\,(x_{1}, \ldots, x_{n})$  for  $ \tilde {h}\,(x_{1}, \ldots, x_{n})$, where
\begin{equation} \label{ash}
 \tilde {h}\,(x_{1}, \ldots, x_{n}) = \vartheta\,  h\,(x_{1}, \ldots, x_{n}).
\end{equation}
Here $\vartheta$ is  infinitely differentiable  function of variable   $P = \sum\limits_{j=1}^{k} p_{k}$  such that  $\vartheta = 1$  if
$P^{2} \geq M^{2}, \: P_{0} > 0; \: \vartheta = 0$,  if $P_{0} \leq 0$.

In order to make the last step it is sufficient to notice that in accordance with spectral properties of   Wightman functions in space-space NC QFT $F_{1} \neq 0$   only  if $P_{0} > 0$ and  $F_{2} \neq 0$   only  if $P_{0} \leq 0$.

Indeed,
$$
W\,(x_{1}, \ldots, x_{k}, x_{k + 1} + \lambda a, \ldots, x_{n} + \lambda a) =
$$
\begin{equation}\label{efonenew }
\left \langle \Psi_{0}, \varphi\,(x_{1}) \ldots   \varphi\,(x_{k})\,U\,(\lambda a)\,\varphi\,(x_{k+1})  \ldots  \varphi\,(x_{n})\Psi_{0} \right\rangle
\end{equation}
and
$$
W\,( x_{k + 1} + \lambda a, \ldots, x_{n} + \lambda a, x_{1}, \ldots, x_{k}) =
$$
\begin{equation}\label{eftwonew}
\left \langle \Psi_{0}, \varphi\,(x_{n}) \ldots   \varphi\,(x_{k + 1})\,U\,(- \lambda a)\,\varphi\,(x_{1})  \ldots  \varphi\,(x_{k})\Psi_{0} \right\rangle.
\end{equation}
So
$$
\int F_{2}\,(x_{1}, \ldots, x_{n})  \tilde {h}\,(x_{1}, \ldots, x_{n})\,d\,x_{1} \ldots d\,x_{n} = 0.
$$
Thus equation (\ref{ef}) is valid also for $F_{1}$ and cluster properties of  Wightman functions in space-space NC QFT are proved.

In order to obtain the stronger result (\ref{cluspropref}) we have to do the calculations similar with ones given in \cite{SW}.

Let us recall that the cluster properties of Wightman functions imply important physical consequences. One of them is the uniqueness of the vacuum state, that is the uniqueness of a translation invariant state.

Let us show that this statement is valid also in space-space NC
QFT. Precisely we show that only one translation invariant state
in respect with commutative coordinates can exist.
In fact, if  there exist two vacuum states  $ \Psi_{0}$ and $ \Psi_{0}^{'} $, we can
always put $ < \Psi_{0},  \Psi_{0}> = 1 , \;   < \Psi_{0}^{'},  \Psi_{0}^{'}> = 1,  \;  < \Psi_{0},  \Psi_{0}^{'}> = 0$.  Then using cluster properties in respect with commutative coordinates, we have
$  < \Psi_{0}^{'},  \Psi_{0}^{'}>  = \lim_{\lambda \to \infty} < \Psi_{0}^{'},  U\,(\lambda\,a)\Psi_{0}^{'}>  =    < \Psi_{0}^{'},  \Psi_{0}> \,   < \Psi_{0},  \Psi_{0}^{'}> = 0$, if  $ a_{0}^{2} -  a_{3}^{2}  = - 1 $.

The proof is completed if  $ \Psi_{0}^{'} $ is a finite linear combination of
vectors \\ $\varphi_{f_1}\, \cdots \,\varphi_{f_n}\,\Psi_{0} $.  If $ \Psi_{0}^{'} $ is an infinite set
of above mentioned vectors, then
\begin{equation} \label{sum}
 \Psi_{0}^{'} = \sum_{0}^{n} \, { c_{k}\,  \varphi_{f_1}\, \cdots \,\varphi_{f_k}\,\Psi_{0} } +  \varepsilon_{n},  \quad  \varepsilon_{n} \to 0, \; \mbox{if} \; n \to \infty.
\end{equation}
As $ U\,(a\,\lambda) \, \Psi'_{0} =     \Psi'_{0} $,  then eq.(\ref{sum}) is valid also for $U\,(a\,\lambda) \, \Psi'_{0}$.
Owing to eqs. (\ref{sum}) and (\ref{clusprop})
$$
<  \Psi'_{0} , U\,(a\,\lambda) \, \Psi'_{0} > = 1 + \delta_{n},  \quad  \delta_{n} \to 0, \; \mbox{if} \; n \to \infty.
$$
Thus we come to the same contradiction as in the first case.

We have proved that cluster properties with respect to the commutative
coordinates lead to the uniqueness of the vacuum state in a similar way as
cluster properties with respect to all coordinates do in the commutative case.
Another important consequence of the cluster properties of Wightman functions,  which is valid in space-space NC QFT, is the statement that
if $\varphi_{f}$ satisfies LCC, but
\begin{equation} \label{loc}
\{\varphi_{f_1}, \varphi_{f_2}^{*}\} = 0,  \quad \{x, y\} = xy + yx,
\end{equation}
then $\varphi_{f} \equiv  0$ \cite{BLT}.
It gives us the possibility to extend the proof of spin-statistic
theorem given in \cite{CMNTV} on complex scalar fields.

 In conclusion, let us show  how cluster properties can be obtained in the NC QFT if LCC is absent. To demonstrate this let us repeat the proof of cluster properties in the book of Strocchi  \cite{St}.
 The only remaining problem is that this proof is valid for usual functions, not for distributions. In order to overcome this difficulty, we have to first consider cluster properties in tubes. Then we use the possibility to go to  zero in the imaginary parts of corresponding variables, and thus extend cluster properties on real variables. Let us point out that as before we have the above mentioned consequence of cluster properties.

\section{Theorems of Reeh and Schlieder in NC QFT}

In the following we shall prove the analogues of the theorems of
Reeh and Schlieder \cite{SW, Jost} for the noncommutative case.

{\bf  Theorem 1} \quad{\it Let supports of functions $\tilde{f_i}$
belong
to $\tilde{O} \times R^2$, where $\tilde{O}$ is any open domain on  variables $x_i^0$ and $x_i^3$. \\
Then there is no vector distinct from zero, which is orthogonal to
all vectors of the type $\varphi_{\tilde{f_1}}\, \cdots
\,\varphi_{\tilde{f_n}}  \, \Psi_0$, supp $\tilde{f_i} \in
\tilde{O} \times R^2$.} First let us consider two vectors
\begin{eqnarray} \label{vec}
\tilde{\Phi}_{n} & = & \varphi_{\tilde{f_1}}\, \cdots
\,\varphi_{\tilde{f_n}}  \, \Psi_0, \quad \mbox{supp} \;
\tilde{f_i} \in
\tilde{O} \times R^2 \quad \forall \; i, \\
\Psi_{m} & = &  \varphi_{f_m}\, \,\ldots \, \varphi_{f_1}\,
\Psi_0.
\end{eqnarray}
On supp ${f_i}$  no restrictions are imposed. We shall prove that
${\Psi}_{m} = 0$, if for any vector $\tilde{\Phi}_{n}$
\begin{equation} \label{17}
\langle {\Psi}_{m}, \tilde{\Phi}_{n} \rangle = 0.
\end{equation}
For the proof it is sufficient to notice that the corresponding
Wightman function
$$
\langle \Psi_0, \varphi\,(y_1)\, \cdots \varphi\,(y_m)  \,\varphi
\,(x_1)\,\, \cdots  \varphi\,(x_n)  \,\Psi_0 \rangle \equiv
W(y_1,...y_m,x_1,...,x_n)
$$
is an analytical function in the variables $- x_1^0 - i\,\eta_0^0,
\; - x_1^3 - i \,\eta_0^3, \; {\nu}_i = {\xi}_i - i \,{\eta}_i, \;
i = 1, \ldots n - 1$, if ${\eta}_i \in V_2^+$. According to the
condition (\ref{17}), this function is equal to zero on the
border, if $x_i \in \tilde{O} \times R^2$. As $\tilde{O}$ is an
open domain, $W(y_1,...y_m,x_1,...,x_n) \equiv 0$. Thus the vector
$\Psi$ is orthogonal to all vectors of the type (\ref{vc}) and,
according to the cyclicity of the vacuum vector, $\Psi_{m} = 0$.
Taking into account that space $J$ is a span of these vectors we
obtain that
\begin{equation} \label{17a}
\langle {\Psi}_{m}, \Psi \rangle  = 0,
\end{equation}
where $\Psi$ is arbitrary. As space $J$ is nondegenerate, this
equality implies that ${\Psi}_{m} = 0$.

To prove the absence of any vector  $\Psi$ orthogonal to all
vectors of the type (\ref{vec}) it is sufficient to notice that
function $\langle \Psi, {\Psi}_{m} \rangle$ is analytical in
$T_{n}^{-}$, and then  use the arguments given above.

Remark that for the proof of the Theorem 1 only the analyticity of the
Wightman functions in the domain $T_n^-$ has been used.

{ \bf Theorem 2} \quad {\it Let the support of $f \in O \times
R^2$, where $O$ is such a domain of commutative variables, for
which domain  $\tilde O \sim O$, satisfying the condition of the
Theorem 1, exists. Then the condition
\begin{equation} \label{18}
\varphi_f  \, \Psi_0 = 0
\end{equation}
implies that
\begin{equation} \label{19}
\varphi_f  \equiv 0,
\end{equation}
if the operator $\varphi_f$ satisfies the LCC.  }

In accordance with LCC
\begin{equation} \label{21}
\varphi_f \,  \, \tilde{\Phi}_{n} = 0,
\end{equation}
if vector $\tilde{\Phi}_{n}$ is defined as in eq. (\ref{vec}). Hence, for any
vector $\Psi$ belonging to the domain of definition of the Hermitian operator
$\varphi_f$,
\begin{equation} \label{22}
\langle \varphi_f  \,   \, \Psi, \tilde{\Phi}_{n} \rangle = \langle \, \Psi,
\varphi_{\bar{f}} \,  \, \tilde{\Phi}_{n} \rangle = 0.
\end{equation}
According to the Theorem 1, the condition (\ref{22}) means that $\varphi_f
\, \Psi = 0$. As the domain of definition of the operator $\varphi_f$ is dense
in $J$, this equality means the validity of the equality (\ref{19}).

{ \bf  Remark} \quad{\it Theorem 2 remains true for any densely
defined operator $\psi_{f}$, mutually local with $\varphi_{\tilde
f}$, i.e. if
\begin{equation} \label{23}
\psi_{f}  \varphi_{\tilde f}  \Phi = \varphi_{\tilde f}  \psi_{f}
\Phi,
\end{equation}
if $supp \; f \in O \times R^2, \; supp\ \tilde f \in \tilde O
\times R^2, \; O \sim \tilde O$ , vector  $\Phi$ belongs to the
domain of definition of operators $\varphi_{\tilde f}$ and
$\psi_{f}$. }

\section{Generalized Haag's Theorem }

Recall the formulation of the generalized Haag's theorem in the
commutative case (\cite{SW}, Theorem~4.17):

{\it Let $\varphi_f^1 \, (t)$ and $\varphi_f^2 \, (t), \, supp\ f \in R^3$ be
two irreducible sets of operators, for which the vacuum vectors $\Psi_0^1$ and
$\Psi_0^2$ are cyclic. Further, let the corresponding Wightman functions be
analytical in the domain $T_n$\footnote{Remark that the required analyticity
of the Wightman functions
follows only from the spectral condition and the $SO (1, 3)$ invariance of the theory.}. \\
Then the two-, three- and four-point Wightman functions coincide
in the two theories if there is a unitary operator $V$, such that}
\begin{eqnarray}
&&1)\ \ \varphi_f^2 \, (t) = V \,\varphi_f^1 \, (t) \, V ^ *,
\label{con1}\\
&&2)\ \ \Psi_0^2 = C \, V \,\Psi_0^1, \quad C \in \mathbb C, \quad
|C | = 1.\label{con2}
\end{eqnarray}

It should be emphasized that actually the condition  $2)$ is a
consequence of condition $1)$ with rather general assumptions (see
the {\bf Statement} below). In the formulation of  Haag's theorem
it is assumed that the formal operators ${\varphi}_i \,(t,
\vec{x}) $ can be smeared only on the spatial variables. This
assumption is natural also in noncommutative case if $\theta^{0i}
=0 $.

Let us consider Haag's theorem in the $SO (1, 1)$  invariant
field theory and show that the corresponding equality is true only
for two-point Wightman functions.

For the proof we first note that in the noncommutative case, just
as in the commutative one, from conditions  $1)$ and  $2)$ it
follows that the Wightman functions in the two theories coincide
at equal times
\begin{equation} \label{32}
\langle \Psi_0^1, \varphi_1 \, (t, \vec{x_1}) \, \tilde \star  \cdots
\varphi_1 \, (t, \vec{x_n}) \, \Psi_0^1 \rangle = \langle \Psi_0^2, \varphi_2
\, (t, \vec{x_1}) \, \tilde \star  \cdots \varphi_2 \, (t, \vec{x_n}) \,
\Psi_0^2 \rangle.
\end{equation}

Having written down the two-point Wightman functions $W_i \, (x_1,
x_2), \; i = 1, 2$ as  $W_i \, (u_1, v_1; u_2, v_2)$, where $u_i =
\{x_i^0, x_i^3 \}, \; v_i = \{x_i^1, x_i^2 \}$ we can write for
them equality (\ref {32}) as:
\begin{equation} \label{32a}
W_1 \, (0, {\xi}^3; v_1, v_2) = W_2 \, (0, {\xi}^3; v_1, v_2),
\end{equation}
where $\xi = u_1 - u_2, \; v_1$ and $v_2$ are arbitrary vectors.
Now we  notice that, due to the $SO (1, 1)$ invariance,
\begin{equation} \label{32b}
W_i \, (0,{\xi}^3; v_1, v_2) = W_i \, (\tilde \xi; v_1, v_2)
\end{equation}
hence,
\begin{equation} \label{32c}
W_1 \, (\tilde \xi; v_1, v_2) = W_2 \, (\tilde \xi; v_1, v_2),
\end{equation}
where $\tilde \xi$ is any Jost point. Due to the analyticity of the Wightman
functions in the commuting variables they are completely determined by their
values at the Jost points. Thus at any $\xi$ from the equality (\ref{32c}), it
follows that
\begin{equation} \label{32d}
W_1 \, (\xi; v_1, v_2) = W_2 \, (\xi; v_1, v_2).
\end{equation}
As $v_1$ and $v_2$ are arbitrary, the formula (\ref{32d}) means
the equality of two-point Wightman functions at all values of
arguments.

Thus, for the equality of the two-point Wightman functions in two
theories related by the conditions (\ref{con1}) and (\ref{con2}),
the $SO (1, 1)$ invariance of the theory and corresponding
spectral condition are sufficient.

It is impossible to extend this proof to three-point Wightman
functions. Indeed, let us write down $W_i \, (x_1, x_2, x_3)$ as
$W_i \, (u_1, u_2, u_3; v_1, v_2, v_3)$, where vectors $u_i$ and
$v_i$ are determined  as before. Equality (\ref{32a}) means that
\begin{equation} \label{32e}
W_1 \, (0, \xi_1^3, 0, \xi_2^3; v_1, v_2, v_3) = W_2 \, (0,
\xi_1^3, 0, \xi_2^3; v_1, v_2, v_3),
\end{equation}
$v_1, v_2, v_3$ are arbitrary. In order to have equality of the
three-point Wightman functions in the two theories from the $SO (1, 1)$ invariance, the existence of transformations $\Lambda
\in SO (1, 1)$ connecting the points $(0, \xi_1^3)$ and $(0,
\xi_2^3)$ with an open vicinity of Jost points is necessary. That
would be possible, if there existed two-dimensional vectors
$\tilde{\xi}_1$ and $\tilde{\xi}_2$, $(\tilde{\xi}_i = \Lambda \,
(0, \xi_i^3))$, satisfying the inequalities:
$$
{(\tilde{\xi}_1)} ^2 < 0, \quad{(\tilde{\xi}_2)} ^2 < 0, \quad |
(\tilde{\xi}_1, \tilde{\xi}_2) | < \sqrt{{(\tilde{\xi}_1)}^2
{(\tilde{\xi}_2)}^2}.
$$
These inequalities are similar to the corresponding inequalities
in the commutative case (see equation (4.87) in \cite{SW}).
However, it is easy to check that the last of these inequalities
can not be fulfilled, while the first two are fulfilled.

Let us show now that the condition (\ref{con2}) actually is a
consequence of the condition (\ref{con1}).

{\bf  Statement} \quad{\it  Condition (\ref{con2}) is fulfilled,
if the vacuum vectors $\Psi_0^i$ are unique, normalized,
translationally invariant vectors with respect to translations
$U_i \, (a)$ along the axis $x^3$.}

It is easy to see that the operator $U_1^{- 1} \, (a) \, V ^{- 1} \, U_2 \,
(a) \, V$ commutes with operators $\varphi_f^1 \,(t)$ and, owing to the
irreducibility of the set of these operators, it is proportional to the
identity operator. Having considered the limit $a = 0$, we see that
\begin{equation} \label{34}
U_1 ^{ - 1} \, (a) \, V ^{- 1} \, U_2 \, (a) \, V = \mathbb I.
\end{equation}
From the equality (\ref{34}) it follows directly that if
\begin{equation} \label{35}
U_1 \, (a) \, \Psi_0^1 = \Psi_0^1,
\end{equation}
then
\begin{equation} \label{36}
U_2 \, (a) \, V \,\Psi_0^1 = V \,\Psi_0^1,
\end{equation}
i.e. the condition (\ref{con2}) is fulfilled. If the theory is
translationally invariant in all variables, the equality
(\ref{36}) is true, if the vacuum vector is unique, normalized,
translationally invariant in the spatial coordinates.

The most important consequence of the generalized  Haag theorem is
the following statement: if one of the two fields related by
conditions (\ref{con1}) and (\ref{con2}) is a free field, the
other is also free. In  deriving this result the equality of the
two-point Wightman functions in the two theories  and LCC are
used. In \cite{CPT} it is proved that this result is valid also in
the noncommutative theory, if $\theta^{0i} = 0$.

Here we obtain the close result in the $SO (1, 1)$ symmetric
theory using the spectral conditions  and translational invariance
only with respect to the commutating coordinates.  In this case
the equality of the two-point Wightman functions in the two
theories leads to the conclusion that if LCC (\ref{wftxi}) is
fulfilled and the current in one of the theories is equal to zero,
for example, $j_f^1 = 0$, then $j_f^2 = 0$ as well; $\; j_f^{i} =
(\square + m^{2}) \, \varphi_f^{i}$. Indeed as $W_{1} \, (x^1,
x^2) = W_{2} \, (x^1, x^2)$,
\begin{equation} \label{43}
< \Psi_0^1,  j^1_{\bar{f}} \,  j^1_{f} \,  \, \Psi_0^1 > = <
\Psi_0^2, j^2_{\bar{f}} \,  j^2_{f}  \, \Psi_0^2> = 0,
\end{equation}
since $j_f^1  = 0$. Hence,
$$
j_f^2  \, \Psi_0^2 = 0.
$$
Here we assume that  $J$ is a positive metric space.  It is
sufficient to take advantage of the Theorem 2 from which follows
that $j_f^2 = 0$ (see the Remark after {\bf Theorem 2}), since LCC
implies mutual local commutativity of a field operator and the
corresponding current.

Let us proceed now to the $SO (1, 3)$ symmetric theory. In
this case we show  that from the equality of the four-point
Wightman functions for the fields ${\varphi}_f^1 \, (t)$ and
${\varphi}_f^2 \, (t)$, related by the conditions (\ref{con1}) and
(\ref{con2}), which takes place in the commutative theory, an
essential physical consequence follows. Namely, for such fields
the elastic scattering amplitudes of the corresponding theories
coincide, and hence, due to the optical theorem, the total
cross-sections coincide as well. In particular, if one of these
fields, for example, ${\varphi}_f^1$ is a trivial field, i.e. the
corresponding $S$ matrix is equal to unity, also the field
${\varphi}_f^2$ is free. In the derivation of this result the
local commutativity condition is not used. The statement follows
directly from the Lehmann-Symanzik-Zimmermann reduction formulas
\cite{LSZ}. Here and below dealing with the commutative case in
order not to complicate formulas we consider operators
${\varphi}_1 \, (x)$ and ${\varphi}_2 \, (x)$ as they are given in
a point.

Let $< p_3, p_4 | p_1, p_2 >_{i}, \; i = 1,2$ be an elastic
scattering amplitudes  for the fields ${\varphi}_1 \, (x)$ and $
{\varphi}_2 \, (x)$ respectively. Owing to the reduction formulas,
$$
< p_3, p_4 | p_1, p_2 >_{i} \, \sim \, \int \,d\,x_1 \cdots d\,x_4
\, e ^{i \, (-p_1 \,x_1 - p_2 \,x_2 + p_3 \,x_3 + p_4 \,x_4)} \,
\cdot
$$
\begin{equation} \label{41}
\prod_{j = 1} ^{4} \, (\square_{j} + m ^{2}) \, < 0 |
T\,\varphi_{i} \, (x_1) \, \cdots \, \varphi_{i} \, (x_4) | 0 >,
\end{equation}
where $T \,\varphi_{i} \, (x_1) \, \cdots \, \varphi_{i} \, (x_4)$
is the chronological product of operators. From the equality
$$
W_2 \, (x_1, \ldots, x_4) = W_1 \, (x_1, \ldots, x_4)
$$
it follows that
\begin{equation} \label{42}
< p_3, p_4 | p_1, p_2 >_{2} = < p_3, p_4 | p_1, p_2 >_{1}
\end{equation}
for any $p_{i}$. Having applied this equality for the forward
elastic scattering amplitudes, we obtain that, according to the
optical theorem, the total cross-sections for the fields
${\varphi}_1 \, (x)$ and ${\varphi}_2 \, (x)$ coincide. If now the
$S$-matrix for the field ${\varphi}_1 \, (x)$ is unity, then it is
also unity for field ${\varphi}_2 \, (x)$. We stress that  the
equality of the four-point Wightman functions in the two theories
related by the conditions (\ref{con1}) and (\ref{con2}) are valid
only in the commutative field theory but not in the noncommutative
case.

\section{Equivalence of various
 conditions of local commutativity in QFT}

Let us show that in the commutative case, when Wightman functions
are analytical ones in the usual domain, the conditions
(\ref{wftx}) and (\ref{wftxi}) are equivalent to the standard
conditions of WLC and LC, i.e. the latter remain valid if the
condition (\ref{pp}) is fulfilled. In effect, (\ref{wftx}) is  a
sufficient condition for the theory to be CPT invariant
\cite{AGM}. However, in the commutative case, from CPT invariance
the standard condition of WLC follows, \cite{SW}--\cite{BLOT}.

The equivalence of LCC (\ref{wftxi}) with the standard one follows
from the fact that, for the validity of usual LCC its validity on
arbitrary small spatially divided domains is sufficient (see \cite
{BLOT}, Proposal 9.12). Indeed, validity of "noncommutative" LCC
(\ref{wftxi}) in the commutative case means validity of standard
LCC in the domain ${(x^{0} - y^{0})}^2 - {(x^{3} - y^{3})}^2 < 0,
\; x^{k}, y^{k}, \; k = 1, 2$ are arbitrary. This domain satisfies
the requirements of the above mentioned statement.

Besides we can replace (\ref{wftxi}) with the formally weaker
condition, requiring that it is valid only when
\begin{equation} \label{lsq}
{\left (x_i^0 - x_j^0 \right)}^2 - {\left (x_i^3 - x_j^3\right)}^2
< - l^{2}, \; \forall \; i, j,
\end{equation}
where $l$  is any fixed fundamental length. Indeed, in the
commutative theory, according to the results of Wightman, Petrina
and Vladimirov (see \cite{Vlad}, Chapter~5 and references therein)
the condition
\begin{equation} \label{45}
[\varphi \, (x), \varphi \, (y)] = 0, \quad (x - y)^{2} < - l^{2},
\end{equation}
for any finite $l$, is equivalent to standard LCC $(l = 0)$.
Similarly, if (\ref{wftxi}) is fulfilled at (\ref{lsq}), then it is
fulfilled also  at $l = 0$.

Thus, the analysis of Wightman functions in NC QFT, carried out in
this and our previous works \cite{CMNTV,CPT,VM05},
shows that the basic axiomatic results are valid (or have
analogues) in NC QFT as well, at least in the case when
$\theta^{0i} = 0$.

\section*{Acknowledgements}
The support of the Academy of Finland under the Projects No.~136539 and No.~140886 is acknowledged.

\end{document}